\documentclass[pra,showpacs,12pt,tightenlines]{revtex4}
\usepackage{amsmath}
\usepackage{amssymb}
\usepackage{dsfont}
\usepackage{graphicx}
\newcommand{\half}{\mbox{$\textstyle \frac{1}{2}$}}

\newcommand{\re}{\mbox{$\rm e$}}

\begin{document}
\title[Coquaternionic quantum dynamics for two-level systems]
{Coquaternionic quantum dynamics for two-level systems}
\author{Dorje C. Brody${}^1$ and Eva-Maria Graefe${}^2$}
\affiliation{${}^1$ Mathematical Sciences, Brunel University, Uxbridge UB8 3PH, UK\\ 
${}^2$ Department of Mathematics, Imperial College London, London SW7 2AZ, UK}

\begin{abstract}
The dynamical aspects of a spin-$\frac{1}{2}$ particle in Hermitian coquaternionic 
quantum theory is investigated. It is shown that the time evolution exhibits three 
different characteristics, depending on the values of the parameters of the 
Hamiltonian. When energy eigenvalues are real, the evolution is either isomorphic 
to that of a complex Hermitian theory on a spherical state space, or else it remains 
unitary along an open orbit on a hyperbolic state space. When energy eigenvalues 
form a complex conjugate pair, the orbit of the time evolution closes again even 
though the state space is hyperbolic. 
\end{abstract}

\maketitle
\vspace{0.4cm}


Over the last decade or so there have been considerable interests in the study of 
complexified dynamical systems; both classically \cite{Bender2,Korsch,Most06,
BHH,BHK,Fring} and quantum mechanically \cite{Bender,Znojil,BBJ,Mostafa,Rotter,
Rivers,Dorey,Witten2010,Graefe3,guenther,Moiseyevbook}. 
For a classical system, its complex extension typically involves the use of 
complex phase-space variables: $(x,p)\to(x_0+i x_1,p_0+ip_1)$. Hence the 
dimensionality of the phase space, i.e. the dynamical degrees of freedom, is 
doubled, and the Hamiltonian $H(x,p)$ in general also becomes complex. For a 
quantum system, on the other, its complex extension typically involves the use of 
a Hamiltonian that is not Hermitian, whereas the dynamical degrees of freedom 
associated with the space of states---the quantum phase space variables---are kept 
real. However, a fully complexified quantum dynamics, analogous to its classical 
counterpart, can be formulated, where state space variables are also 
complexified~\cite{Nesterov,BG}.

The present authors recently observed that there are two natural ways in which 
quantum dynamics can be extended into a fully complex domain~\cite{BG}, where 
both the Hamiltonian \textit{and} the state space are complexified. In short, one is 
to let state space variables and Hamiltonian be quaternion valued; the other is to let 
them coquaternion valued. The former is related to quaternionic quantum mechanics 
of Finkelstein and others \cite{Finkelstein,Adler}, whereas the latter possesses 
spectral structures similar to those of PT-symmetric quantum theory of Bender and 
others \cite{Bender,Znojil,BBJ,Mostafa}. 
The purpose of this paper is to work out in some detail the dynamics of an 
elementary quantum system of a spin-$\frac{1}{2}$ particle under a coquaternionic 
extension, in a manner analogous to the quaternionic case investigated elsewhere 
\cite{BG2}. 

As illustrated in~\cite{BG}, a coquaternionic dynamical system arises from the 
extension of the real and the imaginary parts of the state vector in the complex-$j$ 
direction, where $j$ is the second coquaternionic `imaginary' unit (described below). 
The general dynamics is 
governed by a coquaternionic Hermitian Hamiltonian, whose eigenvalues are either 
real or else appear as complex conjugate pairs. Here we examine the evolution of 
the expectation values of the five Pauli matrices generated by a generic $2\times2$ 
coquaternionic Hermitian Hamiltonian. We shall find that, depending on the values 
of the parameters appearing in the Hamiltonian, the dynamics can be classified into 
three cases: (a) the eigenvalues of $H$ are real and the dynamics is strongly 
unitary in the sense that the `real part' of the dynamics on the reduced state space 
is indistinguishable from that generated by a standard complex Hermitian 
Hamiltonian; (b) the eigenvalues of $H$ are real and the states evolve unitarily into 
infinity without forming closed orbits; and (c) the eigenvalues of $H$ form a 
complex-conjugate pair but the dynamics remains weakly unitary in the sense that 
the real part of the dynamics, although generating closed orbits, no longer lies on 
the state space of a standard complex Hermitian system. Interestingly, properties 
(b) and (c) are in some sense interchanged in a typical PT-symmetric Hamiltonian 
where the orbits of a spin-$\frac{1}{2}$ system are closed when eigenvalues are 
real and open otherwise. These characteristics are related to the three cases 
investigated recently by Kisil \cite{Kisil} in a more general context of Heisenberg 
algebra, based on the use of: (i) spherical imaginary unit $i^2=-1$; (ii) parabolic 
imaginary unit $i^2=0$; and (iii) hyperbolic imaginary unit $i^2=1$. The use 
of coquaternionic Hermitian Hamiltonians thus provides a concise way of 
visualising these different aspects of generalised quantum theory.  

Before we analyse the dynamics, let us begin by briefly reviewing some properties 
of coquaternions that are relevant to the ensuing discussion. 
Coquaternions~\cite{Cockle}, perhaps more commonly known as split quaternions, 
satisfy the algebraic relation 
\begin{eqnarray}
i^2 = -1, \quad j^2 = k^2 = ijk = +1
\end{eqnarray}
and the skew-cyclic relation 
\begin{eqnarray}
ij=-ji=k, \quad jk=-kj=-i, \quad ki=-ik=j .
\end{eqnarray}
The conjugate of a coquaternion $q = q_0 + i q_1 + j q_2 + k q_3$ is ${\bar q} = 
q_0 - i q_1 - j q_2 - k q_3$. It follows that the squared modulus of a coquaternion 
is indefinite: ${\bar q}q = q_0^2 + q_1^2 - q_2^2 - q_3^2$. Unlike quaternions, a 
coquaternion need not have an inverse $q^{-1}={\bar q}/({\bar q}q)$ if it is null, i.e. 
if ${\bar q}q =0$. The polar decomposition of a coquaternion is thus more intricate 
than that of a quaternion. If a coquaternion $q$ has the property that ${\bar q}q>0$ 
and that its imaginary part also has a positive norm so that 
$q_1^2 - q_2^2 - q_3^2>0$, then $q$ can be written in the form 
\begin{eqnarray}
q = |q| \re^{{\boldsymbol i}_q \theta_q} = |q|(\cos\theta_q + 
{\boldsymbol i}_q \sin\theta_q), 
\label{eq:3}
\end{eqnarray}
where 
\begin{eqnarray}
{\boldsymbol i}_q = \frac{iq_1+jq_2+kq_3}{\sqrt{q_1^2-q_2^2-q_3^2}} 
\quad {\rm and} \quad  
\theta_q = \tan^{-1}\left(\frac{\sqrt{q_1^2-q_2^2-q_3^2}}{q_0}\right). 
\end{eqnarray} 
That a coquaternion with `time-like' imaginary part admits the representation 
(\ref{eq:3}) leads to the strong unitary dynamics generated by a coquaternionic 
Hermitian Hamiltonian. On the other hand, if ${\bar q}q>0$ but $q_1^2 - q_2^2 - 
q_3^2<0$, i.e. if the imaginary part of $q$ is `space-like', then 
\begin{eqnarray}
q = |q| \re^{{\boldsymbol i}_q \theta_q} = |q|(\cosh\theta_q + 
{\boldsymbol i}_q \sinh\theta_q),
\end{eqnarray}
where 
\begin{eqnarray}
{\boldsymbol i}_q = \frac{iq_1+jq_2+kq_3}{\sqrt{-q_1^2+q_2^2+q_3^2}} 
\quad {\rm and} \quad 
\theta_q = \tanh^{-1}\left(\frac{\sqrt{-q_1^2+q_2^2+q_3^2}}{|q_0|}\right). 
\end{eqnarray}
If ${\bar q}q>0$ and 
$q_1^2 - q_2^2 - q_3^2=0$, then $q=q_0(1+{\boldsymbol i}_q)$, where 
${\boldsymbol i}_q=q_0^{-1}(iq_1+jq_2+kq_3)$ is the null pure-imaginary 
coquaternion. Finally, if ${\bar q}q<0$, then we have 
\begin{eqnarray}
q = |q| \re^{{\boldsymbol i}_q \theta_q} = |q|(\sinh\theta_q + 
{\boldsymbol i}_q \cosh\theta_q),
\end{eqnarray}
where 
\begin{eqnarray}
{\boldsymbol i}_q = \frac{iq_1+jq_2+kq_3}{\sqrt{-q_1^2+q_2^2+q_3^2}} 
\quad {\rm and} \quad  
\theta_q = \tanh^{-1}\left(\frac{\sqrt{-q_1^2+q_2^2+q_3^2}}{q_0}\right). 
\end{eqnarray}
As indicated above, the fact that the polar decomposition of a coquaternion is 
represented either in terms of trigonometric functions or in terms of hyperbolic 
functions manifest itself in the intricate mixture of spherical and hyperbolic 
geometries associated with the state space of a spin-$\frac{1}{2}$ system, as 
we shall describe in what follows. 

In the case of a coquaternionic matrix ${\hat H}$, its Hermitian conjugate 
${\hat H}^\dagger$ is defined in a manner identical to a complex matrix, i.e. 
${\hat H}^\dagger$ is the coquaternionic conjugate of the transpose of ${\hat H}$. 
Therefore, for a coquaternionic two-level system, a generic Hermitian 
Hamiltonian satisfying ${\hat H}^\dagger={\hat H}$ can be expressed in the 
form 
\begin{eqnarray}
{\hat H} = u_0 {\mathds 1} + \sum_{l=1}^5 u_l {\hat\sigma}_l, 
\label{eq:x9}
\end{eqnarray}
where $\{u_l\}_{l=0..5}\in{\mathds R}$, and 
\begin{eqnarray}
{\hat\sigma}_1 &=& \left( \begin{array}{cc} 0 & 1 \\ 
1 & 0 \end{array} \right), \quad\! 
{\hat\sigma}_2 = \left( \begin{array}{cc} 0 & -i \\ 
i & 0 \end{array} \right), \quad\!
{\hat\sigma}_3 = \left( \begin{array}{cc} 1 & 0 \\ 
0 & -1 \end{array} \right), \nonumber \\ && 
{\hat\sigma}_4 = \left( \begin{array}{cc} 0 & -j \\ 
j & 0 \end{array} \right), \quad\! 
{\hat\sigma}_5 = \left( \begin{array}{cc} 0 & -k \\ 
k & 0 \end{array} \right) 
\end{eqnarray} 
are the coquaternionic Pauli matrices. The eigenvalues of the Hamiltonian 
(\ref{eq:x9}) are given by 
\begin{eqnarray}
E_{\pm} = u_0\pm\sqrt{u_1^2+u_2^2+u_3^2-u_4^2-u_5^2}. 
\end{eqnarray}
Thus, they are both real if $u_1^2+u_2^2+u_3^2>u_4^2+u_5^2$; otherwise they 
form a complex conjugate pair. This, of course, is a characteristic feature of a 
PT-symmetric Hamiltonian. 

A unitary time evolution in a coquaternionic quantum theory is generated by a 
one-parameter family of unitary operators $\re^{-{\hat A}t}$, where ${\hat A}$ is 
skew-Hermitian: ${\hat A}^\dagger=-{\hat A}$. As in the case of complex quantum 
theory, we would like to let the Hamiltonian ${\hat H}$ be the generator of the 
dynamics. For this purpose, let us write 
\begin{eqnarray}
{\boldsymbol i} = \frac{1}{\nu}\, (iu_2+ju_4+ku_5), 
\end{eqnarray}
where $\nu=\sqrt{u_2^2-u_4^2-u_5^2}$ if $u_4^2+u_5^2<u_2^2$, and 
$\nu=\sqrt{u_4^2+u_5^2-u_2^2}$ if $u_2^2<u_4^2+u_5^2$. Then we set 
${\hat A}={\boldsymbol i} {\hat H}$ and the Schr\"odinger equation in units 
$\hbar=1$ is thus given by (cf. \cite{BG2})
\begin{eqnarray}
|{\dot \psi}\rangle = - {\boldsymbol i} {\hat H} |\psi\rangle.
\end{eqnarray}
It is worth remarking that when $u_4^2+u_5^2<u_2^2$ we have ${\boldsymbol i}^2
=-1$, whereas when $u_2^2<u_4^2+u_5^2$ we have ${\boldsymbol i}^2=+1$. In 
either case ${\boldsymbol i} {\hat H}$ is a skew-Hermitian operator satisfying 
$({\boldsymbol i} {\hat H})^\dagger=-{\boldsymbol i} {\hat H}$; thus 
$\re^{-{\boldsymbol i} {\hat H}t}$ formally generates a unitary time evolution that 
preserves the norm $\langle\psi|\psi\rangle={\bar\psi}_1\psi_1+{\bar\psi_2}\psi_2$, 
where ${\bar\psi}$ is the coquaternionic conjugate of $\psi$ so that $\langle\psi|$ 
represents the Hermitian conjugate of $|\psi\rangle$. 
The conservation of the norm can be checked directly by use of the explicit form 
of the Schr\"odinger equation in terms of the components of the state vector: 
\begin{eqnarray}
\left( \begin{array}{c} {\dot\psi}_1 \\ {\dot\psi}_2 \end{array} \right) = 
\left( \begin{array}{c} 
-(u_0+u_3) {\boldsymbol i} \psi_1 - u_1 {\boldsymbol i} \psi_2 - \nu \psi_2 \\ 
-(u_0-u_3) {\boldsymbol i} \psi_2 - u_1 {\boldsymbol i} \psi_1 + \nu \psi_1 
\end{array} \right) .
\label{eq:11}
\end{eqnarray} 
Here we have assumed $u_4^2+u_5^2<u_2^2$ so that $\nu=
\sqrt{u_2^2-u_4^2-u_5^2}$; if $u_2^2<u_4^2+u_5^2$, we have 
$\nu=\sqrt{u_4^2+u_5^2-u_2^2}$ and the sign of $\nu$ in (\ref{eq:11}) 
changes. 

To investigate properties of the unitary dynamics generated by the Hamiltonian 
(\ref{eq:x9}) we shall derive the evolution equation satisfied by what one might 
call a `coquaternionic Bloch vector' ${\vec\sigma}$, whose components are given 
by 
\begin{eqnarray}
\sigma_l = \frac{\langle\psi|{\hat\sigma}_l|\psi\rangle}{\langle\psi|
\psi\rangle}, \quad l=1,\ldots,5.
\end{eqnarray}
By differentiating $\sigma_l$ in $t$ for each $l$ and using the dynamical equation 
(\ref{eq:11}), we deduce, after rearrangements of terms, the following set of 
generalised Bloch equations: 
\begin{eqnarray}
\half {\dot\sigma}_1 &=& \nu \sigma_3 - \frac{u_3}{\nu} (u_2\sigma_2 
+ u_4\sigma_4 + u_5 \sigma_5) \nonumber \\ 
\half {\dot\sigma}_2 &=& \frac{1}{\nu} (u_2u_3\sigma_1 - u_1u_2 
\sigma_3 + u_0u_5\sigma_4 - u_0u_4 \sigma_5) \nonumber \\ 
\half {\dot\sigma}_3 &=& -\nu \sigma_1 + \frac{u_1}{\nu} (u_2\sigma_2 
+ u_4\sigma_4 + u_5 \sigma_5) \label{eq:x16} \\ 
\half {\dot\sigma}_4 &=& \frac{1}{\nu} (-u_3u_4\sigma_1 + u_0u_5 
\sigma_2 + u_1u_4\sigma_3 + u_0u_2 \sigma_5) \nonumber \\ 
\half {\dot\sigma}_5 &=& \frac{1}{\nu} (-u_3u_5\sigma_1 - u_0u_4 
\sigma_2 + u_1u_5\sigma_3 - u_0u_2 \sigma_4), \nonumber 
\end{eqnarray}
where we have assumed $u_4^2+u_5^2<u_2^2$ so that $\nu=
\sqrt{u_2^2-u_4^2-u_5^2}$. This is the region in the parameter space where the 
coquaternion appearing in the Hamiltonian has a time-like imaginary part. 
Note that these evolution equations preserve the condition: 
\begin{eqnarray}
\sigma_1^2 + \sigma_2^2 + \sigma_3^2 - \sigma_4^2 - \sigma_5^2 = 1,
\label{eq:x21}
\end{eqnarray}
which can be viewed as the defining equation for the hyperbolic state space of a 
coquaternionic two-level system. 

\begin{figure}[t]
\begin{center}
  \includegraphics[scale=0.65]{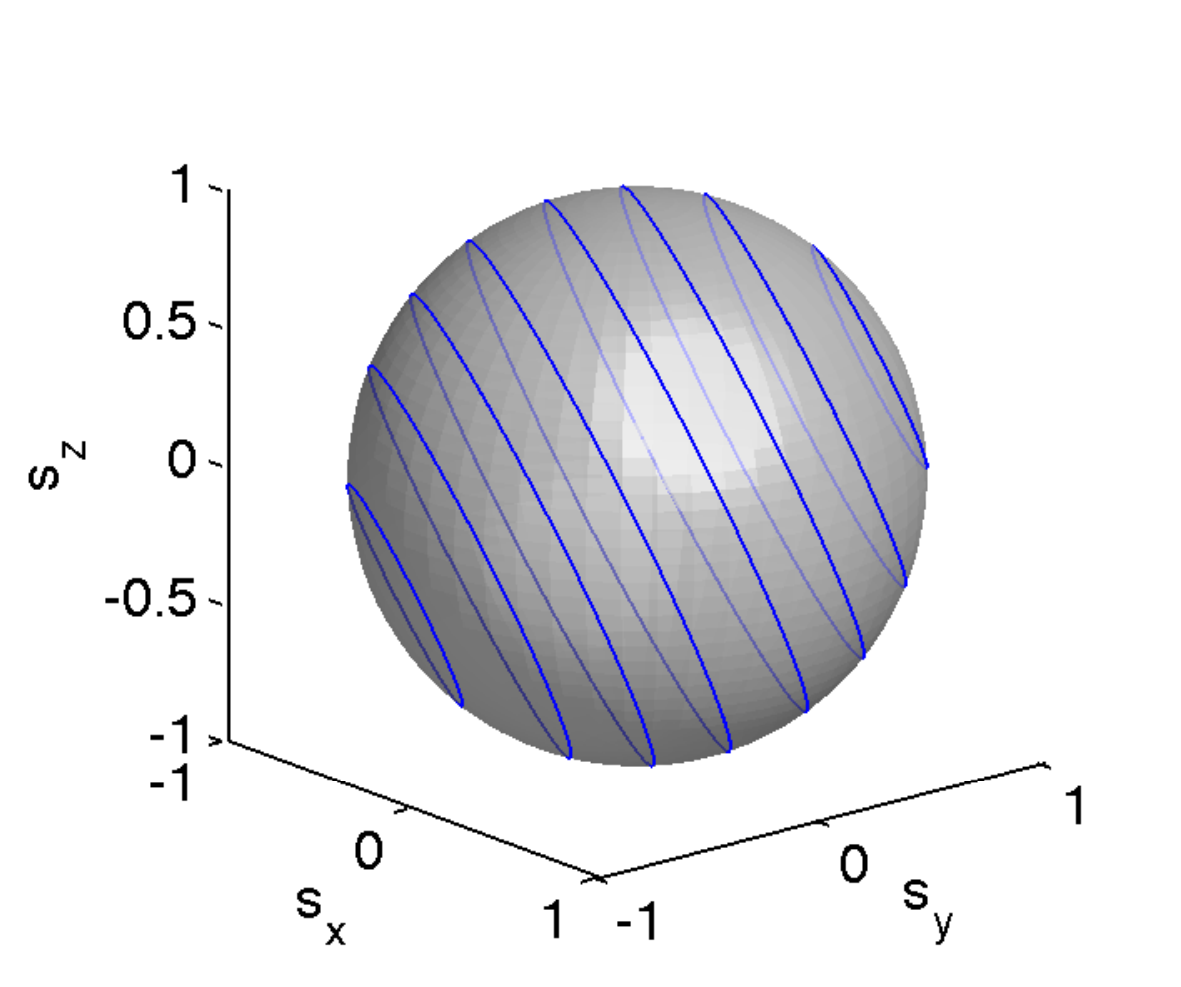}
  \includegraphics[scale=0.65]{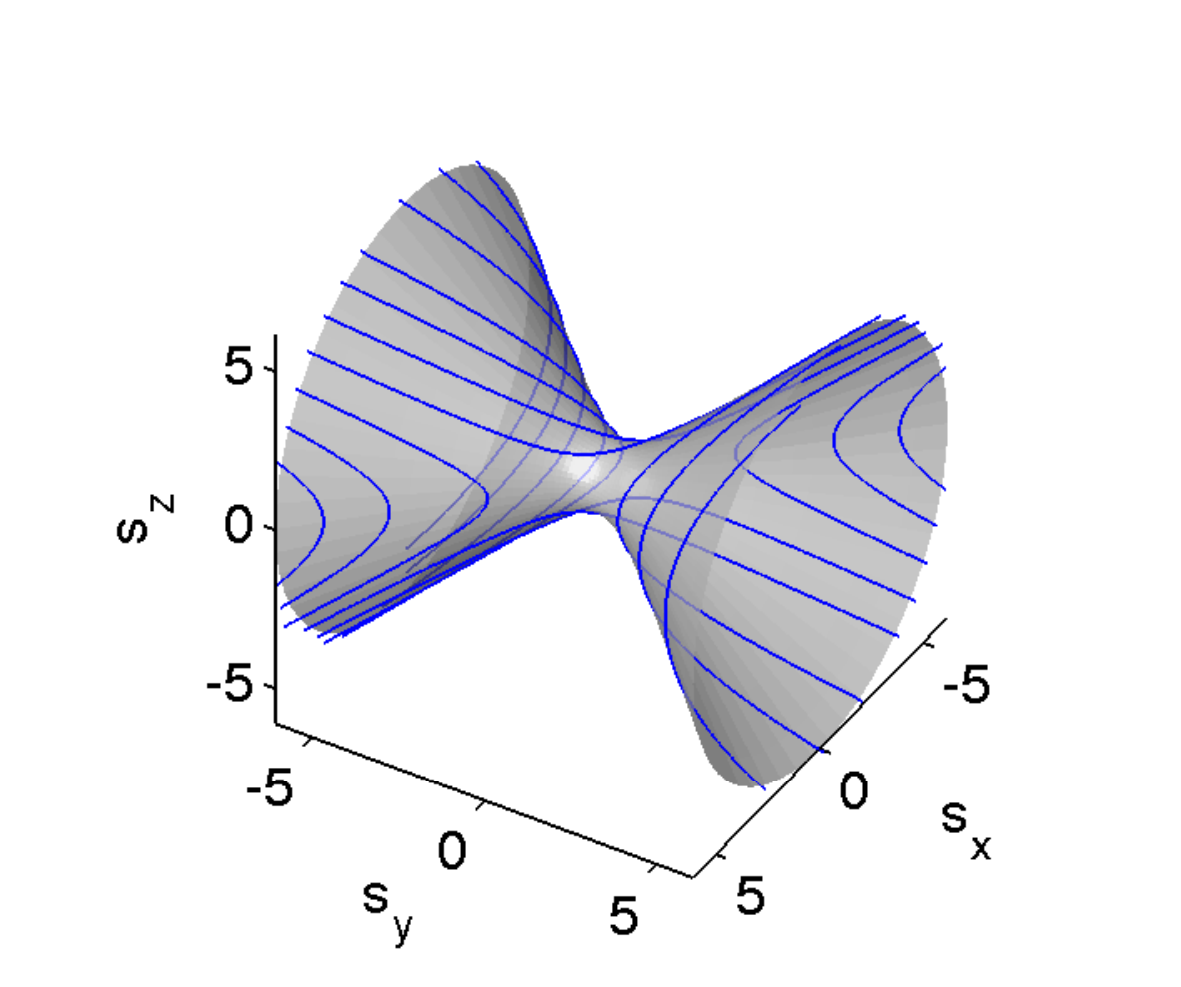}
  \caption{(colour online) 
  \textit{Dynamical trajectories on the reduced state spaces}. 
  In the parameter region $u_2^2>u_4^2+u_5^2$ the reduced state space is 
  just a two-sphere, upon which the dynamical equations (\ref{eq:16}) generate 
  Rabi oscillations (left figure). In the parameter region $u_2^2<u_4^2+u_5^2$ the 
  reduced state space is a two-dimensional hyperboloid, and the dynamical 
  equations (\ref{eq:x25}) generate open trajectories on this hyperbolic state 
  space, if the energy eigenvalues are real (right figure). If the 
  eigenvalues are complex, the open trajectories are rotated to form hyperbolic 
  Rabi oscillations. 
  \label{fig:1} 
  }
\end{center}
\end{figure}

Let us now show how the dynamics can be reduced to three-dimensions so as 
to provide a more intuitive understanding. For this purpose, we define the three 
reduced spin variables
\begin{eqnarray}
\sigma_x = \sigma_1, \quad \sigma_y = \frac{1}{\nu}(u_2\sigma_2+
u_4\sigma_4+u_5\sigma_5), \quad \sigma_z = \sigma_3. 
\label{eq:15}
\end{eqnarray}
We can think of the space spanned by these reduced spin variables as 
representing the `real part' of the state space (\ref{eq:x21}). Then a short 
calculation making use of (\ref{eq:x16}) shows that  
\begin{eqnarray}
\half {\dot\sigma}_x &=& \nu \sigma_z -u_3\sigma_y \nonumber \\ 
\half {\dot\sigma}_y &=& u_3\sigma_x - u_1\sigma_z \label{eq:16} \\ 
\half {\dot\sigma}_z &=& u_1 \sigma_y - \nu \sigma_x,  \nonumber 
\end{eqnarray}
or, more concisely, ${\dot{\vec{\sigma}}}=2{\vec B}\times{\vec\sigma}$ where 
${\vec B}=(u_1,\nu,u_3)$. Hence although the state space of a coquaternionic 
spin-$\frac{1}{2}$ system is a hyperboloid (\ref{eq:x21}), remarkably in the 
region $u_4^2+u_5^2<u_2^2$ the reduced spin variables $\sigma_x,\sigma_y, 
\sigma_z$ defined by (\ref{eq:15}) obey the standard Bloch equations (\ref{eq:16}). 
In particular, the reduced motions are confined to the two sphere $S^2$: 
\begin{eqnarray}
\sigma_x^2+\sigma_y^2+\sigma_z^2={\rm const}., \label{eq:17} 
\end{eqnarray}
where the value of the right side of (\ref{eq:17}) depends on the initial condition 
(but is positive and is time independent). 
Put the matter differently, in the parameter region $u_4^2+u_5^2<u_2^2$, 
the dynamics on the reduced state space $S^2$ induced by a coquaternionic 
Hermitian Hamiltonian is indistinguishable from the conventional unitary 
dynamics generated by a complex Hermitian Hamiltonian. This corresponds to 
the situation in a PT-symmetric quantum theory whereby in some regions of 
the parameter space the Hamiltonian is complex Hermitian (e.g., a harmonic 
oscillator in the Bender-Boettcher Hamiltonian family $H=p^2+x^2(ix)^\epsilon$ 
\cite{Bender}, or the six-parameter $2\times2$ matrix family in \cite{wang}). 
Some examples of dynamical trajectories are sketched in figure~\ref{fig:1}. 

The evolution of the other dynamical variables $\sigma_2,\sigma_4,\sigma_5$ 
can be analysed as follows. Recall that the dynamics (\ref{eq:16}) preserves the 
relation (\ref{eq:17}). Thus, by subtracting (\ref{eq:17}) from (\ref{eq:x21}) and 
rearranging terms we deduce that 
\begin{eqnarray}
-(u_2\sigma_4+u_4\sigma_2)^2 + (u_4\sigma_5-u_5\sigma_4)^2 - 
(u_5\sigma_2+u_2\sigma_5)^2 = {\rm const}. \label{eq:18}
\end{eqnarray}
This shows that the evolution of the vector $(\sigma_2,\sigma_4,\sigma_5)$ is 
confined to a hyperbolic cylinder. 
It turns out that the time evolution of these `hidden' dynamical variables 
$\sigma_2,\sigma_4,\sigma_5$ can also be represented in a form similar to 
Bloch equations if we transform the variables according to $\sigma_{y_1}=
u_4\sigma_5-u_5\sigma_4$, $\sigma_{y_2}=u_5\sigma_2+u_2\sigma_5$, and 
$\sigma_{y_3}=u_2\sigma_4+u_4\sigma_2$. In terms of these auxiliary variables 
we have 
\begin{eqnarray}
\half {\dot\sigma}_{y_1} &=& -\frac{u_0}{\nu} (u_5\sigma_{y_2} +u_4
\sigma_{y_3})  \nonumber \\ 
\half {\dot\sigma}_{y_2} &=& -\frac{u_0}{\nu} (u_2\sigma_{y_3} + u_5
\sigma_{y_1}) \label{eq:x22} \\ 
\half {\dot\sigma}_{y_3} &=& -\frac{u_0}{\nu} (u_4 \sigma_{y_1} - u_2 
\sigma_{y_2}).  \nonumber 
\end{eqnarray}
It should be evident that these dynamics are confined to a hyperboloid: 
\begin{eqnarray}
-\sigma_{y_1}^2+\sigma_{y_2}^2+\sigma_{y_3}^2={\rm const}. 
\end{eqnarray} 
Note, however, that when $u_0=0$ we have ${\dot\sigma}_{y_1} = 
{\dot\sigma}_{y_2} = {\dot\sigma}_{y_3} =0$ from (\ref{eq:x22}), while 
$\sigma_2,\sigma_4,\sigma_5$ are in general evolving in time. Hence in 
transforming the variables into $\sigma_{y_1},\sigma_{y_2},\sigma_{y_3}$, 
part of the information concerning the dynamics is lost. 

We see from (\ref{eq:17}) and (\ref{eq:18}) that on the `imaginary part' of the state 
space the dynamics is endowed with hyperbolic characteristics, which nevertheless 
is not visible on the reduced state space, or the `real part' of the state space $S^2$ 
spanned by $\sigma_x,\sigma_y, \sigma_z$. 

When $u_2^2=u_4^2+u_5^2$ so that the imaginary part of the coquaternion 
appearing in the Hamiltonian is null, a calculation shows that the reduced spin 
variables obey the following dynamical equations: 
\begin{eqnarray}
\half {\dot\sigma}_x &=& -u_3\sigma_y \nonumber \\ 
\half {\dot\sigma}_y &=& -u_3\sigma_x + u_1\sigma_z \label{eq:16.5} \\ 
\half {\dot\sigma}_z &=& u_1 \sigma_y ,  \nonumber 
\end{eqnarray}
and preserve $\sigma_x^2-\sigma_y^2+\sigma_z^2$. 

\begin{figure}[t]
\begin{center}
  \includegraphics[scale=0.65]{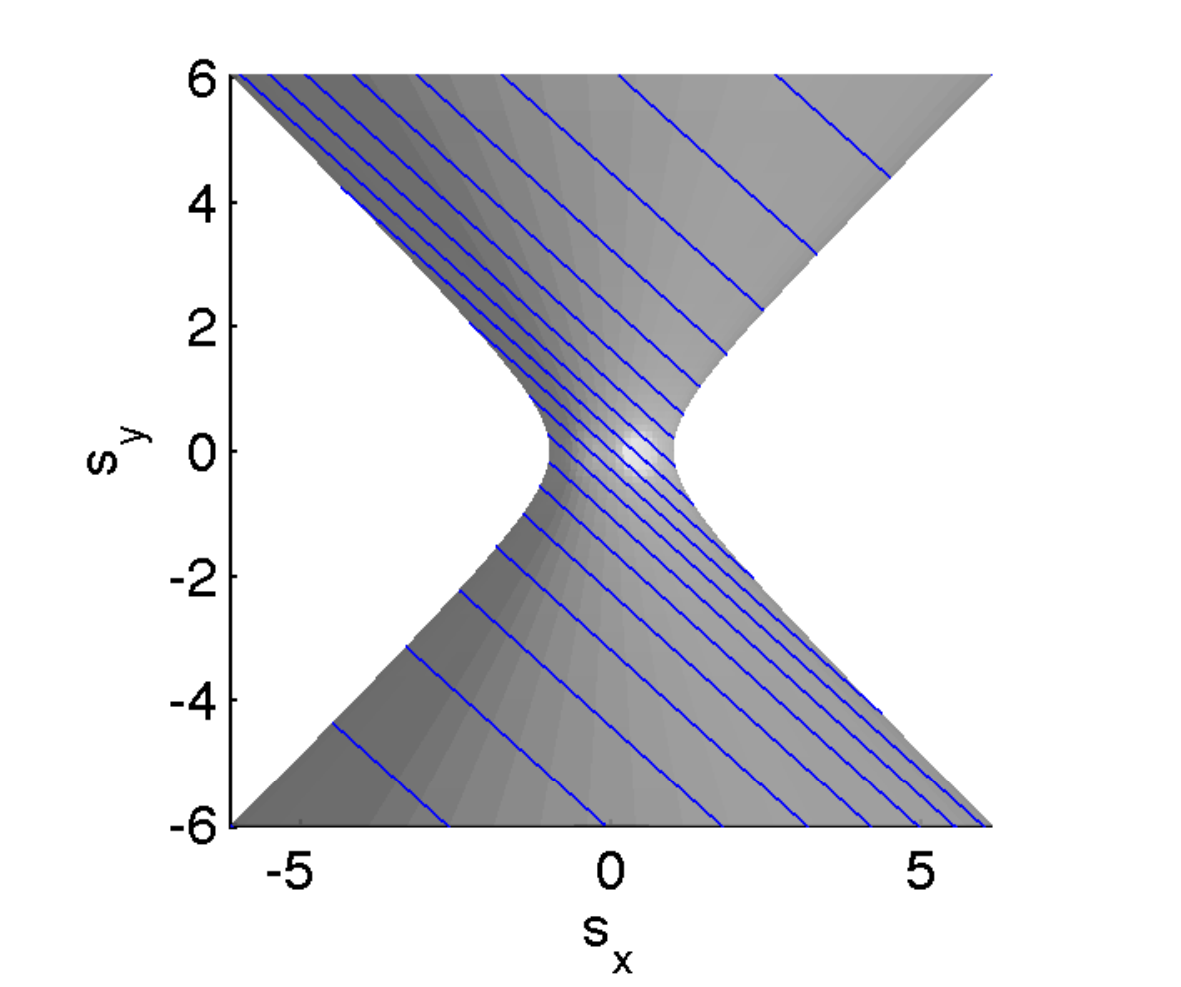}
  \includegraphics[scale=0.65]{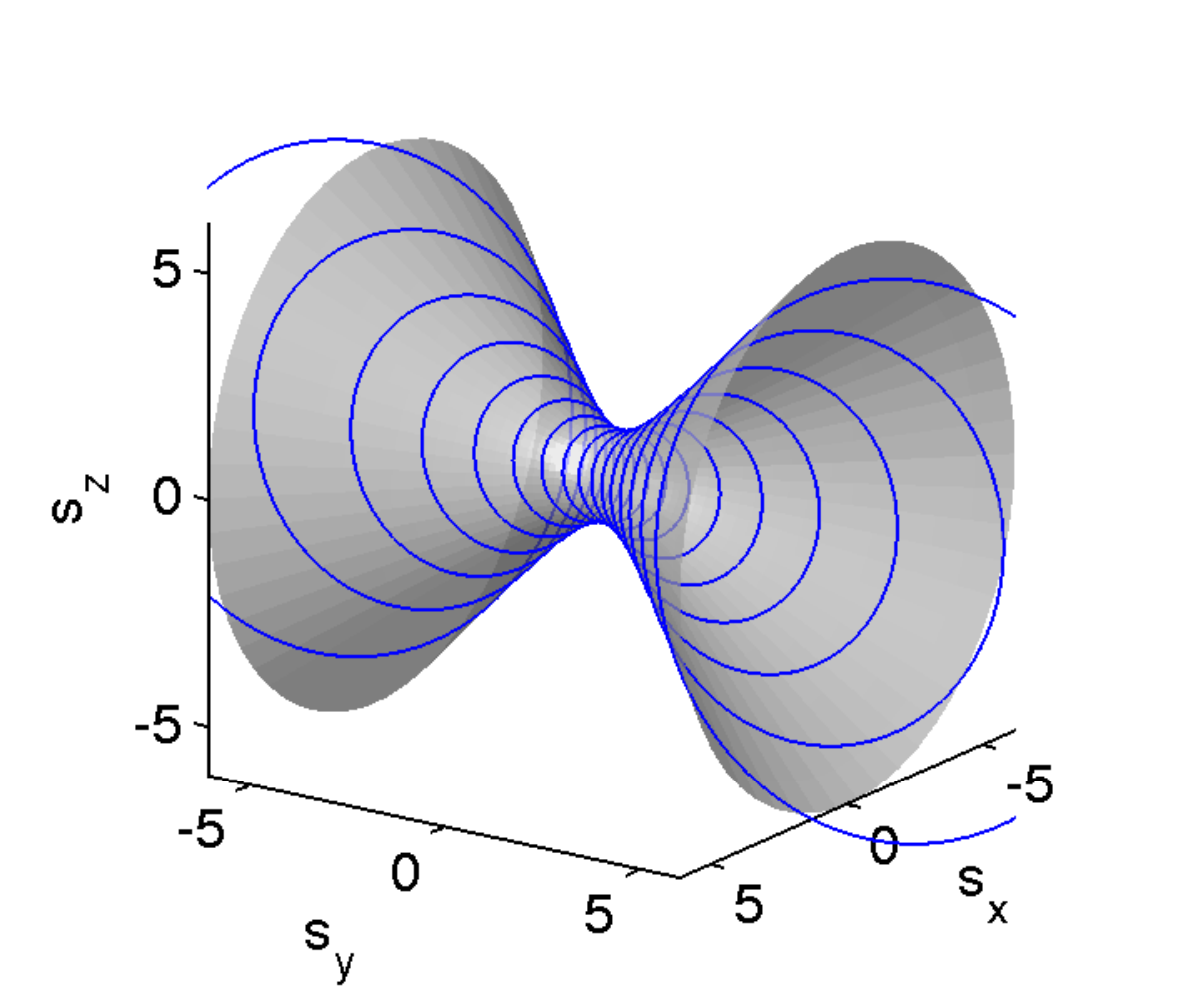}
  \caption{(colour online) 
  \textit{Conic sections and PT phase transition: changes of orbit structures}. 
  A projection of the orbits on the hyperboloid, for parameters just above the 
  transition to complex energy eigenvalues, is shown on the left side. The 
  orbits form circular sections. On the right side we plot orbits of hyperbolic 
  Rabi oscillations further into complex energy eigenvalues. The energy 
  eigenvalues determine the angle between the axis of rotation and the axis 
  of the hyperboloid. When eigenvalues are complex, the axis of rotation is 
  within the hyperboloid, leading to closed orbits on the state space 
  generated by circular sections. When the imaginary part of the coquaternion 
  appearing in the Hamiltonian is null, we have parabolic sections of the 
  hyperboloid; whereas when the energy eigenvalues are real, the angle of the 
  two axes is larger than $\pi/4$, and open orbits are generated by hyperbolic 
  sections. 
  \label{fig:2} 
  }
\end{center}
\end{figure}

When $u_2^2<u_4^2+u_5^2$ so that the imaginary part of the coquaternion in 
the Hamiltonian is space-like, the structure of the state space, as well as the 
dynamics, change, and they exhibit an interesting and nontrivial behaviour. The 
five-dimensional spin variables in this case evolve according to 
\begin{eqnarray}
\half {\dot\sigma}_1 &=& -\nu \sigma_3 - \frac{u_3}{\nu} (u_2\sigma_2 
+ u_4\sigma_4 + u_5 \sigma_5) \nonumber \\ 
\half {\dot\sigma}_2 &=& \frac{1}{\nu} (u_2u_3\sigma_1 - u_1u_2 
\sigma_3 + u_0u_5\sigma_4 - u_0u_4 \sigma_5) \nonumber \\ 
\half {\dot\sigma}_3 &=& \nu \sigma_1 + \frac{u_1}{\nu} (u_2\sigma_2 
+ u_4\sigma_4 + u_5 \sigma_5) \\ 
\half {\dot\sigma}_4 &=& \frac{1}{\nu} (-u_3u_4\sigma_1 + u_0u_5 
\sigma_2 + u_1u_4\sigma_3 + u_0u_2 \sigma_5) \nonumber \\ 
\half {\dot\sigma}_5 &=& \frac{1}{\nu} (-u_3u_5\sigma_1 - u_0u_4 
\sigma_2 + u_1u_5\sigma_3 - u_0u_2 \sigma_4), \nonumber 
\end{eqnarray}
where $\nu=\sqrt{u_4^2+u_5^2-u_2^2}$. 
These evolution equations preserve the normalisation (\ref{eq:x21}). However, 
in the region $u_2^2<u_4^2+u_5^2$ the reduced spin variables $\sigma_x,
\sigma_y,\sigma_z$ defined by (\ref{eq:15}) no longer obey the standard Bloch 
equations (\ref{eq:16}); instead, they satisfy
\begin{eqnarray}
\half {\dot\sigma}_x &=& -\nu \sigma_z -u_3\sigma_y \nonumber \\ 
\half {\dot\sigma}_y &=& -u_3\sigma_x + u_1\sigma_z \label{eq:x25} \\
\half {\dot\sigma}_z &=& u_1 \sigma_y + \nu \sigma_x,  \nonumber 
\end{eqnarray}
and preserve the relation 
\begin{eqnarray}
\sigma_x^2-\sigma_y^2+\sigma_z^2 = {\rm const}. 
\label{eq:22}
\end{eqnarray}
We thus see that at the level of reduced spin variables in three dimensions, 
the state space changes from a two-sphere (\ref{eq:17}) to a hyperboloid 
(\ref{eq:22}), as the parameters $u_2,u_4,u_5$ appearing in the 
Hamiltonian change. This transition corresponds to the transition from a complex 
Hermitian Hamiltonian into a PT-symmetric non-Hermitian Hamiltonian. Since 
the energy eigenvalues can still be real even when $u_2^2<u_4^2+u_5^2$, 
we expect the dynamics to exhibit two distinct characteristics depending on 
whether the reality condition $u_1^2+u_2^2+u_3^2>u_4^2+u_5^2$ is satisfied. 
Indeed, we found that on a hyperbolic state space, orbits of the unitary dynamics 
associated with real energies are the ones that are open and run off to infinities. 
Conversely, when the reality condition is violated, these open orbits are in effect 
Wick rotated to generate closed orbits. These features can be identified by a 
closer inspection on the structure of the underlying state space, upon which the 
dynamical orbits lie. In particular, (\ref{eq:x25}) shows that the dynamics generates 
a rotation around the axis $(u_1,\nu,u_3)$; whereas the state space (\ref{eq:22}) is 
a hyperboloid about the axis $(0,1,0)$. We have sketched in figure~\ref{fig:2}  
dynamical orbits resulting from (\ref{eq:x25}), indicating that there indeed is a 
transition from open to closed orbits as real eigenvalues turn into complex 
conjugate pairs. 

Intuitively, one might have expected an opposite transition since in a 
PT-symmetric model of a spin-$\frac{1}{2}$ system the renormalised Bloch 
vectors on a spherical state space follow closed orbits when eigenvalues 
are real, whereas sinks and sources are created when eigenvalues become 
complex \cite{Graefe3}. The apparent opposite behaviour seen here is presumably 
to do with the facts that the underlying state space is hyperbolic, not spherical, 
and that no renormalisation is performed here. In figure~\ref{fig:2} we have 
sketched some dynamical trajectories when energy eigenvalues are complex. 
A projection of the dynamical orbits from the $\sigma_z$ axis (for the choice of 
parameters used in these plots) shows in which way the topology of the orbits 
are affected by the reality of the energy eigenvalues. 

The evolutions of the other dynamical variables $\sigma_2,\sigma_4,\sigma_5$ 
are confined to the space characterised by the relation 
\begin{eqnarray}
(u_2\sigma_4+u_4\sigma_2)^2 - (u_4\sigma_5-u_5\sigma_4)^2 + 
(u_5\sigma_2+u_2\sigma_5)^2 = {\rm const}., \label{eq:23}
\end{eqnarray}
instead of the relation (\ref{eq:18}) of the previous case. However, if we define, 
as before, three auxiliary variables $\sigma_{y_1}=u_4\sigma_5-u_5\sigma_4$, 
$\sigma_{y_2}=u_5\sigma_2+u_2\sigma_5$, and $\sigma_{y_3}=u_2\sigma_4 
+u_4\sigma_2$, then the dynamical equations satisfied by these variables are 
identical to those in (\ref{eq:x22}), except, of course, that the definition of $\nu$ 
is different. 

It is interesting to remark that when the imaginary part of the coquaternion 
appearing in the Hamiltonian is space-like, the imaginary unit ${\boldsymbol i}$ 
has the characteristic of a `double number' or a `Study number' introduced by 
Clifford \cite{Clifford}, that is, ${\boldsymbol i}^2=1$. Quantum theories generated 
by such a number field (instead of the field of complex numbers) and other 
hyperbolic generalisations, as well as various issues that might arise from such 
generalisations, have been discussed by various authors (e.g., 
\cite{Hudson,kocik1999}; see also \cite{Kisil} and references cited therein). 
The use of coquaternionic Hermitian Hamiltonian thus captures dynamical 
behaviours of different generalisations of quantum mechanics in a simple 
unified scheme.

\vskip 8pt We thank the participants of the international 
conference on Analytic and Algebraic Methods in Physics VII, Prague,  
2011, for stimulating discussions. EMG is supported by an Imperial College 
Junior Research Fellowship.





\begin{thebibliography}{999}




\bibitem{Bender2} Bender,~C.~M., Boettcher~S. \& Meisinger,~P.~N. 
1999 PT-symmetric quantum mechanics. {\em J. Math. Phys.} 
\textbf{40}, 2201. 

\bibitem{Korsch}
Kaushal,~R.~S. \& Korsch,~H.~J. 2000 Some remarks on complex 
Hamiltonian systems. {\em Phys. Lett.} A\textbf{276}, 47.

\bibitem{Most06}
Mostafazadeh,~A. 2006 Real description of classical Hamiltonian 
dynamics generated by a complex potential. {\em Phys. Lett.} A{\bf 357}, 
177.

\bibitem{BHH} 
Bender,~C.~M., Holm,~D.~D. \& Hook,~D.~W. 2007 
Complex trajectories of a simple pendulum. 
{\em J. Phys.} A\textbf{40}, F81. 

\bibitem{BHK} 
Bender,~C.~M., Hook,~D.~W. \& Kooner,~K.~S. 2010 
Classical particle in a complex elliptic pendulum. 
{\em J. Phys.} A\textbf{43}, 165201. 

\bibitem{Fring} 
Cavaglia,~A., Fring,~A. \& Bagchi,~B. 2011
PT-symmetry breaking in complex nonlinear wave equations and their 
deformations. arXiv:1103.1832

\bibitem{Bender} 
Bender,~C.~M. \& Boettcher~S. 1998 
Real spectra in non-Hermitian Hamiltonians having PT symmetry. 
{\em Phys. Rev. Lett.} \textbf{80}, 5243.  

\bibitem{Znojil} 
L\'evai,~G. \& Znojil,~M. 2000 
Systematic search for PT-symmetric potentials with real energy spectra. 
{\em J. Phys.} A\textbf{33}, 7165. 

\bibitem{BBJ} Bender,~C.~M., Brody,~D.~C. \& Jones,~H.~F. 2002 
Complex extension of quantum mechanics. 
{\em Phys. Rev. Lett.} \textbf{89}, 270401. 

\bibitem{Mostafa} Mostafazadeh,~A. 2002 Pseudo-Hermiticity 
versus PT-symmetry III: Equivalence of pseudo-Hermiticity and 
the presence of antilinear symmetries. {\em J. Math. Phys.} 
\textbf{43}, 3944. 

\bibitem{Rotter}
J. Oko{\l}owicz,~J., P{\l}oszajczak,~M. \& Rotter,~I. 2003
Dynamics of quantum systems embedded in a continuum. 
{\em Phys. Rep.} \textbf{374}, 271. 

\bibitem{Rivers}
Jones,~H.~F. \& Rivers,~R.~J. 2009 
Which Green functions does the path integral for quasi-Hermitian 
Hamiltonians represent? 
{\em Phys. Lett.} A\textbf{373}, 3304. 

\bibitem{Dorey} 
Dorey,~P., Dunning,~C., Lishman,~A. \& Tateo,~R. 2009 
PT symmetry breaking and exceptional points for a class of inhomogeneous 
complex potentials. 
{\em J. Phys.} A\textbf{42}, 465302. 

\bibitem{Witten2010} 
Witten,~E. 2010 
A new look at the path integral of quantum mechanics. 
arXiv:1009.6032. 

\bibitem{Graefe3} Graefe,~E.~M., Korsch,~H.~J. \& 
Niederle,~A.~E. 2010 Quantum-classical correspondence for 
a non-Hermitian Bose-Hubbard dimer. {\em Phys. Rev.} 
A\textbf{82}, 013629. 

\bibitem{guenther} 
G\"unther,~U. \& Kuzhel,~S. 2010 
PTÐsymmetry, Cartan decompositions, Lie triple systems and Krein
space-related Clifford algebras. {\em J. Phys.} A\textbf{43}, 39002. 

\bibitem{Moiseyevbook}
Moiseyev,~N. 2011 \textit{Non-Hermitian Quantum Mechanics}, 
(Cambridge: Cambridge University Press).

\bibitem{Nesterov} Nesterov,~A.~I. 2009 Non-Hermitian quantum 
systems and time-optimal quantum evolution. {\em SIGMA} 
\textbf{5}, 069. 

\bibitem{BG} Brody,~D.~C. \& Graefe,~E.~M. 2011 
On complexified mechanics and coquaternions. {\em J. Phys.} 
A\textbf{44}, 072001. 

\bibitem{Finkelstein} Finkelstein,~D., Jaueh,~J.~M., Schiminovieh,~S. 
\& Speiser,~D. 1962 Foundations of quaternion quantum mechanics. 
{\em J. Math. Phys.} \textbf{3}, 207. 

\bibitem{Adler} Adler,~S.~L. 1995 {\em Quaternionic Quantum 
Mechanics and Quantum Fields}. (Oxford: Oxford University 
Press). 

\bibitem{BG2} Brody,~D.~C. \& Graefe,~E.~M. 2011  
Six-dimensional space-time from quaternionic quantum mechanics. 
arXiv:1105.3604. 

\bibitem{Kisil} 
Kisil,~V.~V. 2010
Erlangen programme at large 3.1: Hypercomplex representations of the 
Heisenberg group and mechanics. arXiv:1005.5057v2 

\bibitem{Cockle} Cockle,~J. 1849 On systems of algebra 
involving more than one imaginary. {\em Phil. Magazine} 
\textbf{35}, 434.  

\bibitem{wang} Wang,~Q., Chia,~S., \& Zhang,~J. 2010 PT 
symmetry as a generalisation of Hermiticity. {\em J. Phys.} 
A\textbf{43}, 295301. 

\bibitem{Clifford}
Clifford,~W.~K. 1878
Applications of Grassmann's extensive algebra. 
{\em Am. J. Math.} \textbf{1}, 350.

\bibitem{Hudson}
Hudson,~R.~L. 1966 
Generalised translation-invariant mechanics. 
D.~Phil. Thesis, Bodleian Library, Oxford.

\bibitem{kocik1999} 
Kocik,~J. 1999 
Duplex numbers, diffusion systems, and generalized quantum mechanics. 
{\em Int. J. Theo. Phys.} \textbf{38}, 2221. 




\end{thebibliography}
\end{document}